\documentclass[
    aps,
    pra,
    longbibliography,
    superscriptaddress,amsmath,amssymb,amsfonts, 
    tightenlines,
    twocolumn,
    ]{revtex4-2}

\usepackage[utf8]{inputenc}  
\usepackage[T1]{fontenc}     
\usepackage[english]{babel}  

\usepackage[colorlinks = true, linkcolor=teal, citecolor=teal, urlcolor  = teal]{hyperref}

\usepackage{graphicx} 
\usepackage[babel]{microtype}  
\usepackage{amsmath,amssymb,amsthm,bm,amsfonts,mathrsfs,bbm} 
\usepackage{setspace}
\usepackage{times}
\usepackage{braket}
\usepackage[bottom]{footmisc}
\usepackage{blochsphere}
\usepackage{pgfplots}
\usepackage{amsthm}

\pgfplotsset{compat=1.17}
\usetikzlibrary{calc,3d,shapes, pgfplots.external, intersections}
\usepackage{algorithm}
\usepackage{physics}
\usepackage[noend]{algorithmic}
\usepackage{eqparbox}

\usepackage{tikz}
\usepackage{mathtools}
\usetikzlibrary{arrows.meta}
\usepackage{csquotes}

\usepackage{xspace}  
\usepackage{pgfplots}
\usepackage{verbatim}
\usepackage{amsmath}

\newtheorem{theorem}{Theorem}

\newtheoremstyle{note}
{3pt}
{3pt}
{}
{}
{\itshape}
{:}
{.5em}
{}

\newcommand\id{\leavevmode\hbox{\small1\kern-3.3pt\normalsize1}}

\newcommand{\x}{{\bf x}}

\newtheorem{conjecture}[theorem]{Conjecture}
\newtheorem{corollary}[theorem]{Corollary}

\newtheorem{definition}[theorem]{Definition}

\definecolor{jens}{rgb}{0.1,0.4,0.6}

\usepackage{pdfpages}
\usepackage{pgffor}
\makeatletter
\AtBeginDocument{\let\LS@rot\@undefined}
\makeatother

\begin{document}


\title{Gauge symmetry permits extremely low-depth random unitary circuits}


\author{Yunfei Wang}
\email{yunfeizl@umd.edu}
\affiliation{Joint Center for Quantum Information and Computer Science, NIST/University of Maryland, College Park, MD 20742, USA}
\affiliation{Department of Physics, University of Maryland, College Park, Maryland 20742, USA}
\affiliation{Joint Quantum Institute, University of Maryland and NIST, College Park, Maryland 20742, USA}


\author{Yuqing Li}
\affiliation{Department of Computer Science, The University of Pittsburgh, Pittsburgh, PA 15260, USA}

\author{Junqi Wang}
\affiliation{Department of Computer Science, The University of Pittsburgh, Pittsburgh, PA 15260, USA}

\author{Alexey V. Gorshkov}
\email{gorshkov@umd.edu}
\affiliation{Joint Center for Quantum Information and Computer Science, NIST/University of Maryland, College Park, MD 20742, USA}
\affiliation{Joint Quantum Institute, University of Maryland and NIST, College Park, Maryland 20742, USA}

\author{Junyu Liu}
\email{junyuliu@pitt.edu}
\affiliation{Department of Computer Science, The University of Pittsburgh, Pittsburgh, PA 15260, USA}

\maketitle

{\bf Continuous global symmetries can obstruct logarithmic-depth symmetric unitary \(k\)-designs and pseudorandom unitaries (PRUs) through long-range correlations. We ask whether gauge symmetries impose analogous limitations. For \(\mathrm{U}(1)\) lattice gauge theories, we identify a randomization scale \(c\), the minimum patch size needed for matter to screen every allowed boundary-flux mismatch, and a diffusive scale governing boundary-flux correlations. Glued circuits built from gauge-symmetric unitaries with subdiffusive support are efficiently distinguishable from gauge-symmetric Haar-random unitaries, whereas local controlled designs can be glued into global approximate designs when overlaps exceed the diffusive scale multiplied by a logarithmic factor. Our analysis suggests incomplete screening limits design guarantees to constant order, whereas strictly locally random systems with complete local screening (\(c=1\)) admit a local tensor-product description, permitting superpolynomial design orders and logarithmic-depth designs and PRUs. We construct a toy model of charged black-hole evaporation in which insufficient screening makes uninterrupted complete discharge exponentially unlikely in \(c\), whereas strict local randomization removes this screening obstruction. This motivates a quantum-information analogue of the weak gravity conjecture, requiring the available charged spectrum to support complete local screening in addition to superextremality. We conjecture that this condition allows gauge-symmetric systems to scramble as rapidly as unconstrained ones and support this by constructing logarithmic-depth approximate strong designs and strong PRUs.}


\section{{Introduction}} \label{sec:introduction}

Random unitary dynamics provides an operational language for quantum scrambling. Approximate unitary \(k\)-designs~\cite{schuster2025randomunitariesextremelylow,LaRacuente_2026,Brand_o_2016,chen2024incompressibilityspectralgapsrandom} and pseudorandom unitaries (PRUs)~\cite{schuster2025randomunitariesextremelylow,ma2025constructrandomunitaries,cui2025unitarydesignsnearlyoptimal,schuster2025strongrandomunitariesfast} formalize when a circuit becomes indistinguishable from Haar-random evolution, and have become central tools in quantum information~\cite{Huang_2020Predicting,Hayden_2007}, many-body physics and quantum cryptography~\cite{Ji_2018,metger2024simple}. In unconstrained many-body systems, a powerful route to such randomness is provided by gluing constructions: independently randomizing overlapping subsystems in two staggered layers can already generate approximate unitary \(k\)-designs at logarithmic depth~\cite{schuster2025randomunitariesextremelylow,schuster2025hardnessrecognizingphasesmatter}. Continuous global symmetries, however, obstruct shallow gluing altogether. An additive conserved charge remains locally detectable under sublinear-depth symmetric dynamics, yielding a distinguishing protocol that rules out such circuits from reproducing globally Haar-random symmetric evolution~\cite{haah2025shortremarksshallowunitary}.

Unlike global symmetries, gauge symmetries represent local redundancies, and gauge-invariant dynamics are restricted to the Gauss-law-constrained physical Hilbert space \(\mathcal H_{\mathrm{phys}}\)~\cite{Kogut:1974ag,Coleman:2018mew,Schwartz:2014sze}. Gauss's law correlates local gauge and matter configurations and can leave some boundary-flux sectors anomalously sparse, thereby obstructing rapid randomization. Crucially, however, these constraints can be screened by charged matter: sufficiently rich local charge spectra can restore an abundance of physical configurations across flux sectors.

Therefore, in this Article, we ask when an Abelian U(1) lattice gauge system can randomize as efficiently as an unconstrained many-body system. We let $n$ denote the total number of lattice sites and use it as the asymptotic parameter. We introduce a diffusive scale $\mathrm{diff}(n)$, beyond which different boundary-flux sectors share the same leading growth rate in their numbers of physical configurations, while their remaining boundary-flux dependence factorizes up to exponentially small corrections. Here the $n$-dependence is an asymptotic convention. In general, the diffusive scale is determined by the local charge and flux spectra; for simplicity, we use $n$ as the single large parameter and allow the corresponding microscopic cutoffs to scale with system size. This lets us express the resulting screening scale as $\mathrm{diff}(n)$ and compare it directly with circuit-depth scaling. Below this scale, residual boundary-flux memory survives local Haar randomization, yielding a distinguishing protocol and a no-go result for gluing constructions built from constituent unitaries with sub-diffusive support. 

Beyond this scale, boundary-flux correlations decay, allowing efficient gluing for sufficiently large patches. We then prove a gauge-symmetric gluing theorem directly on the Gauss-law-constrained physical Hilbert space and obtain low-depth approximate unitary $k$-designs, with a sufficient patch size of order $\mathrm{diff}(n)\log n$ for fixed design order and inverse-polynomial accuracy. When lattice gauge system is strictly locally random, the additional gauge-sector corrections disappear, allowing logarithmic-depth constructions of approximate strong designs and strong PRUs. Gauge-symmetric random dynamics therefore occupy an intermediate regime between globally symmetric and unconstrained systems: Gauss's law introduces a characteristic, model-dependent spatial scale for randomization rather than necessarily enforcing irreducible system-wide memory.

The same screening structure motivates a toy model of charged black-hole evaporation. A ``gauge-charged black hole'' is represented by a shrinking lattice gauge system whose interior is repeatedly randomized within a fixed-charge physical sector, while ``Hawking emission'' is charge blind apart from local gauge constraints. If the available matter spectrum cannot sufficiently screen the residual flux, the system can enter an extremal regime in which uninterrupted complete discharge requires a highly constrained sequence of charged emissions, with probability exponentially suppressed in the randomization scale \(c(n)\). By contrast, complete local screening (\(c=1\)) eliminates this many-body screening obstruction down to the smallest semiclassically resolved configuration.

In its simplest electric form, the Weak Gravity Conjecture posits that a consistent theory of quantum gravity contains charged states with charge-to-mass ratios at least as large as that of an extremal black hole~\cite{Arkani_Hamed_2007,pal2010weakgravityconjecturecentral,Cheung_2014,Harlow_2016,cottrell2017weakgravityconjectureextremal,Hebecker_2017,Andriolo_2018,Cheung_2018,Harlow_2023}. The underlying intuition is that charged black holes should be able to discharge rather than persist as stable charged remnants. Motivated by this discharge principle, we propose a lattice-screening weak gravity conjecture that supplements the energetic discharge condition with a local screening requirement on the available matter-charge spectrum. In our toy model, this screening requirement coincides with strict local randomization and also removes the additional gauge-induced limitations on randomization. This correspondence motivates a gauge-charged extension of the fast-scrambling conjecture~\cite{Hayden_2007,Sekino_2008}, linking efficient discharge to rapid scrambling of gauge-invariant information. We provide operational evidence for this connection by constructing low-depth gauge-symmetric strong unitary designs and strong PRUs in strictly locally random systems~\cite{schuster2025strongrandomunitariesfast}.

\section{{Results}}

We now present our main results on when gauge-symmetric systems obstruct, or permit, approximate Haar randomness at shallow circuit depth. For concreteness, we consider a \(\mathrm U(1)\) lattice gauge theory on a one-dimensional open chain of \(n\) matter sites with \(n+1\) left-to-right oriented links, including one dangling boundary link at each end~\cite{Brower_1999}. Matter degrees of freedom live on sites $i \in\{1, \ldots, n\}$ and carry integer charges $q_i \in\left\{q_{-}, q_{-}+1, \ldots, q_{+}\right\}$, while the gauge degrees of freedom live on the links and are labeled by non-negative electric fluxes $E_{i, i+1} \in\left\{0,1, \ldots, E_{\max }\right\}$. These finite ranges specify the allowed local charge and flux spectra of our model and may be understood as those retained in an effective low-energy description. The resulting kinematical Hilbert space is $\mathcal H_{\mathrm{kin}} = \left(\mathbb C^{q_+-q_-+1}\right)^{\otimes n} \otimes \left(\mathbb C^{E_{\max}+1}\right)^{\otimes(n+1)}$. Although we use the one-dimensional open chain to make the structure transparent, the definitions and constructions below extend naturally to more general lattice geometries, building on the gluing framework of refs.~\cite{schuster2025randomunitariesextremelylow,schuster2025hardnessrecognizingphasesmatter,schuster2025strongrandomunitariesfast}.

The physical Hilbert space $\mathcal{H}_{\text {phys}} \subset \mathcal{H}_{\text {kin}}$ is obtained by imposing Gauss's law at every matter site. With our orientation convention, the local gauge generator is $Q_i=q_i+E_{i-1, i}-E_{i, i+1}$, and the physical Hilbert space is defined as
\begin{equation}
    \mathcal H_{\mathrm{phys}} = \big\{ \ket{\psi}\in\mathcal H_{\mathrm{kin}}:~Q_i\ket{\psi}=0, \forall i \in \{1,\ldots,n\} \big\}~.
\end{equation}
Thus physical states are invariant under local gauge transformations. Moreover, once the electric fluxes on all links are specified, Gauss's law uniquely determines the matter charge at each site, $q_i=E_{i, i+1}-E_{i-1, i}$. The physical configurations can therefore be parametrized entirely by their link-flux configurations, subject to the requirement that the induced matter charges lie within the allowed local charge range.

\subsection{The randomization and diffusive scale of Abelian lattice gauge theories}

To characterize how efficiently a gauge-symmetric system can erase correlations among its gauge degrees of freedom, we first introduce the randomization scale of the system. We begin by defining the randomization over a subsystem as follows.
\begin{definition}[Randomization on a patch, 1D lattice]
\label{def:RandomizationScale}
We say that a connected patch \(\xi\) is randomizable if every boundary-flux configuration (i.e.~sector) permitted by the boundary link Hilbert spaces is realized by at least one physical state on \(\xi\). 
\end{definition}

Let $\xi$ denote a connected patch of sites, and define its oriented boundary-flux difference as: $F_\xi \equiv E_{\partial\xi_{\text{in}}} -  E_{\partial\xi_{\text{out}}}~,$ where $\partial\xi_{\mathrm{in}}$ and $\partial\xi_{\mathrm{out}}$
denote the incoming and outgoing boundary links immediately
outside the patch $\xi$, respectively. In this 1D system, the maximum and minimum of $F_\xi$ are precisely $\pm E_{\max}$.

\begin{theorem}[Randomization condition, 1D]
\label{thm:LocalRandomization}
A connected-subsystem \(\xi\) is randomizable, if and only if 
\begin{equation} \label{eq:LocalRandomizationConditionMain}
    E_{\max} \le |\xi| q_+~ \quad \text{and} \quad -E_{\max} \ge |\xi| q_-~,
\end{equation}
where $|\xi|$ is the number of sites in $\xi$. The \emph{randomization scale} \(c(n)\) of the system is the smallest integer such that every connected patch \(\xi\) with \(|\xi|\ge c(n)\) satisfies Eqn.~\eqref{eq:LocalRandomizationConditionMain}. If $c = 1$, we call this system \emph{strictly locally random}, in which all gauge degrees of freedoms are independent, and the physical Hilbert space is isomorphic to a local tensor product space: $\mathcal{H}_{\text{phys}} \cong (\mathbb{C}^{E_{\max}+1})^{\otimes (n+1)}$.
\end{theorem}

For example, when the matter-charge spectrum is $q_i \in\{-1,0,1\}$, each site can change the electric flux by at most one unit. Screening the largest possible flux mismatch therefore requires $E_{\max }$ sites, giving $c=E_{\max }$. In the asymptotic convention, \(E_{\max}\) may itself scale with the system size, in which case \(c(n)=E_{\max}(n)\). More generally, the randomization scale is controlled by the ratio between the maximal flux mismatch and the amount of charge available locally to screen it. In this sense, $c(n)$ measures the minimum distance over which arbitrary flux differences can be erased by matter degrees of freedom. We will use the $q_i \in\{-1,0,1\}$ example repeatedly because it makes this interpretation particularly transparent. For more general lattice gauge systems, $c(n)$ is defined as the minimum patch size beyond which every connected region is randomizable, and can depend on both the local charge and flux spectra and the lattice geometry. The detailed proof of Theorem~\ref{thm:LocalRandomization} and explicit construction for higher-dimensional lattices are presented in the Supplementary Material Section~I.B.

A patch larger than the randomization scale $c(n)$ can realize every allowed boundary-flux sector, but those sectors need not yet be statistically equivalent. Consider a connected patch $\xi$ with incoming and outgoing boundary fluxes $E_{\partial \xi_{\text{in}}}=a$ and $E_{\partial \xi_{\text {out}}}=b$, and let $\mathcal{H}_{a, b}^{(\xi)}$ denote the corresponding physical Hilbert space, with \(D_{|\xi|}(a,b)\) its dimension. For $q_i \in\{-1,0,1\}$ and $|\xi|=c(n)$, the extremal sector $\mathcal{H}_{0, E_{\max}}^{(\xi)}$ contains only a single physical configuration, $D_{|\xi|}\left(0, E_{\max}\right)=1$, whereas a neutral sector such as $\mathcal{H}_{0,0}^{(\xi)}$ has dimension exponential in $c(n)$. Thus, although all boundary sectors are kinematically accessible, their dimensions can remain highly uneven: fixing boundary fluxes with a large mismatch strongly constrains the allowed configurations across the patch. This raises a second question: over what length scale is this residual boundary-flux memory finally washed out?

For $q_i \in\{-1,0,1\}$, we find that this occurs only on a larger scale, $\operatorname{diff}(n) \sim c^2(n)=E_{\max }^2(n)$, with $|\xi| \gg$ $\operatorname{diff}(n)$ denoting $|\xi|=\omega(\operatorname{diff}(n))$ as $n \rightarrow \infty$. In this regime, the dimensions $D_{|\xi|}(a,b)$ of different boundary-flux sectors share the same leading growth rate, differing only by boundary-dependent factors:
\begin{equation} \label{eq:DiffScaleDimMain}
    D_{|\xi|\gg\mathrm{diff}(n)}(a,b) = D_{|\xi|\gg\mathrm{diff}(n)}(a',b')\times c(n)^{\mathcal{O}(1)}~,
\end{equation}
for all $a,b,a',b' \in \{0,1,\ldots,E_{\max}\}$. More precisely, these dimensions share the same leading dependence on the patch size $|\xi|$, while their dependence on the two boundary fluxes factorizes:
\begin{equation} \label{eq:Factorization}
    D_{|\xi|}(a,b)
    = \lambda_1^{|\xi|}\phi_1(a)\phi_1(b)
    \left[1+\mathcal{O}\left(e^{-\Omega(|\xi|/\mathrm{diff}(n))}\right)\right]~,
\end{equation}
where $\lambda_1$ determines the common growth with patch size, and $\phi_1(a)$ captures the dependence on each boundary flux. Physically, the influence of a fixed boundary flux has by then diffused across the patch and no longer strongly constrains the opposite boundary. We therefore identify $\mathrm{diff}(n)$ as the characteristic scale for erasing boundary-flux correlations. The detailed calculation and its extension to general Abelian lattice gauge theories on arbitrary lattice geometries are given in Supplementary Information, Section~I.C.2.

\begin{figure*}[!ht]
\centering
\includegraphics[width=0.95\linewidth]{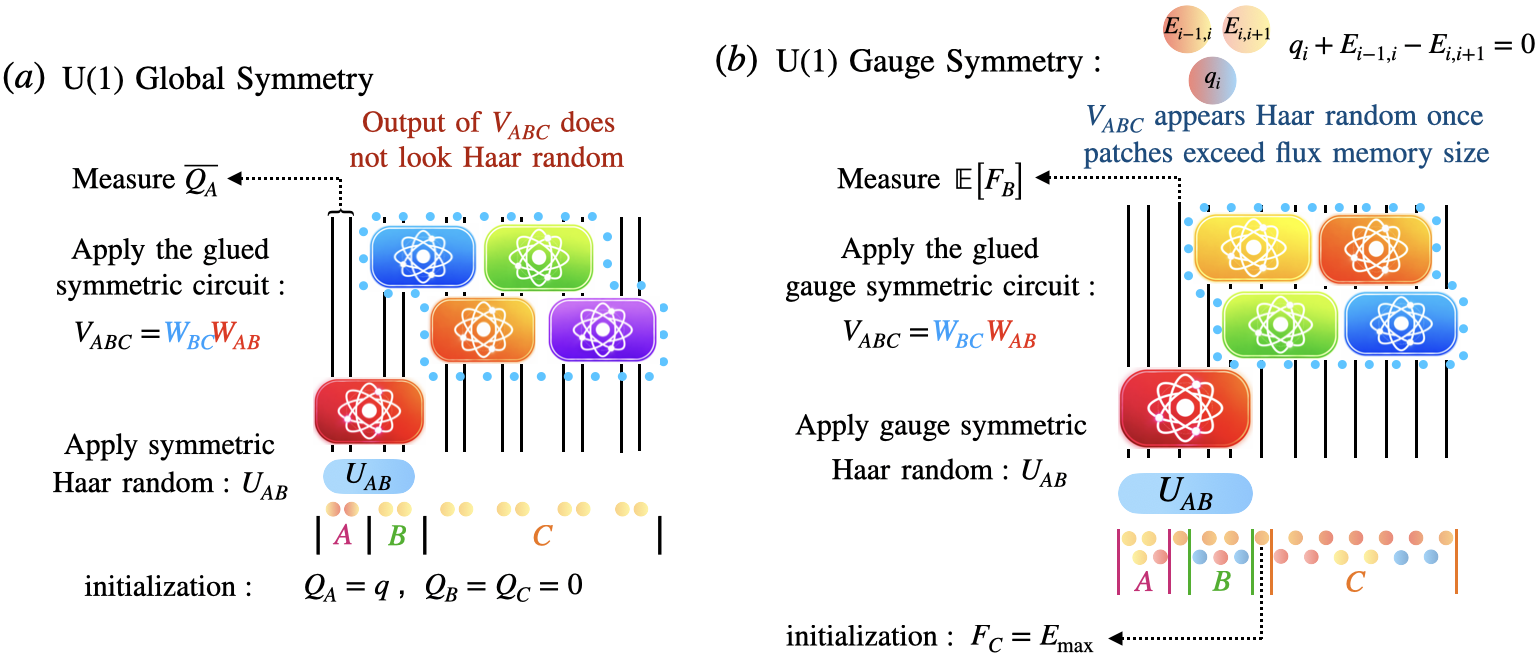}
\caption{A glued circuit \(V_{ABC}=W_{BC}W_{AB}\) built from overlapping globally or gauge-symmetric random unitaries. In both panels, \(W_{AB}\), shown as the leftmost red block, acts on \(AB\), while \(W_{BC}\), indicated by the blue dotted outline, acts on \(BC\). The overlap region \(B\) is shown in green at the bottom; \(A\), of equal size, is shown in pink; and \(C\), representing the remainder of the system, is shown in orange. (a) Distinguishing protocol for a globally conserved charge \cite{haah2025shortremarksshallowunitary}: residual charge remains detectable in \(A\) after the glued evolution. (b) Gauge-symmetric protocol: the upper and lower blobs at the bottom denote gauge and matter degrees of freedom of the system, respectively (see legend at the top of the figure). For sub-diffusive-sized patches, memory of the initialized boundary flux persists in \(F_B\), allowing the glued ensemble to be distinguished from gauge-symmetric Haar-random results.}
\label{fig:NoGoProt}
\centering
\end{figure*}

\subsection{No gluing subdiffusive-supported gauge-symmetric random unitaries}

The obstruction to shallow symmetric gluing \cite{schuster2025randomunitariesextremelylow,LaRacuente_2026} for continuous global symmetries can be understood operationally through the distinguishing protocol \cite{haah2025shortremarksshallowunitary}, illustrated in Figure~\ref{fig:NoGoProt}(a). Unlike unconstrained dynamics, a continuous symmetry protects a commuting conserved charge that cannot be hidden by operator spreading. To expose this residual information, one initializes total charge \(q\) inside a region \(A\) of sublinear size, chooses an adjacent region \(B\) of equal size, and denotes the remainder by \(C\). After symmetric Haar randomization on \(AB\), the expected charge in \(A\) is \(\overline{Q_A}=q/2\). A glued circuit \(V_{ABC}=W_{BC}W_{AB}\) cannot further reduce this expectation: \(W_{BC}\) acts trivially on \(A\), while \(W_{AB}\) is statistically absorbed into the preceding Haar randomization. By contrast, a symmetric Haar-random unitary on the full system spreads the conserved charge throughout \(ABC\), giving \(\overline{Q_A}=q|A|/n\), where $n$ is the total system size. Measuring \(Q_A\) therefore provides a direct distinguishing signal between the glued shallow ensemble and symmetric Haar-random dynamics.

In order to study whether a shallow gauge-symmetric circuit assembled by gluing together random unitaries acting on overlapping local patches can reproduce the statistics of a globally Haar-random gauge-symmetric unitary, we construct a similar protocol which shows that this is impossible when the constituent unitaries have subdiffusive support. We illustrate this explicitly for a one-dimensional open chain with \(q_i\in\{-1,0,1\}\) and \(E_{i,i+1}\in\{0,1,\ldots,E_{\max}\}\), for which the randomization scale is \(c(n)=E_{\max}(n)\) and the diffusive scale is \(\mathrm{diff}(n)=c^2(n)\). The key idea is to initialize a subdiffusive patch in a boundary-flux sector of anomalously small dimension. Specifically, we divide the chain into three consecutive regions $A,B,C$, with the fluxes at the $A$–$B$ and $B$–$C$ interfaces denoted by $F_B$ and $F_C$, respectively. We take region $B$ to have size $|B|\equiv\xi(n)$ and initialize the interface flux between $B$ and $C$ to its maximal value, $F_C=E_{\max}$; see Figure~\ref{fig:NoGoProt}(b).

Gauge-symmetric Haar randomization on patch \(\xi\) is block diagonal in its boundary-flux sectors, $U_\xi = \bigoplus_{a,b} U^{(\xi)}_{a,b}$, where $U^{(\xi)}_{a,b}\in \mathrm{U}\left(D_{|\xi|}(a,b)\right)$. Here for patch $AB$, the unitary $U_{AB} = \bigoplus_{F_C} U^{(AB)}_{F_C}$ is only controlled on the interface flux $F_C$: $U^{(AB)}_{F_C}\in \mathrm{U}\left(D_{\xi(n)}(F_C)\right)$, whereas $U_{BC}$ is only controlled on $F_{B}$. Given this structure, after the initialization, we apply a gauge-symmetric Haar-random unitary \(U_{AB}\), and subsequently apply either the glued circuit \(V_{ABC}=W_{BC}W_{AB}\) or a gauge-symmetric Haar-random unitary \(U_{ABC}\) on the full system. The initialized boundary flux is unchanged by \(U_{AB}\), since the corresponding link lies outside its support. Moreover, \(W_{AB}\) is statistically absorbed into the preceding Haar randomization, while \(W_{BC}\) cannot act on the fluxes outside of $BC$, so \(F_{B}\) is unchanged.

For subdiffusive-sized overlapping patch, $\xi(n) \ll c^2(n)$, fixing the outgoing boundary flux to its maximal value strongly biases the incoming-flux sectors of $A B$ : configurations with large $F_{B}$ are far more numerous than those with small $F_{B}$. For $q_i \in\{-1,0,1\}$,
\begin{equation}\label{eq:ExpecValueE01SmallMain}
\mathbb{E}_{V_{A B C}}\left[F_{B}\right] \approx E_{\max }-\sqrt{\frac{\pi \xi(n)}{3}}+\mathcal{O}(1)~,
\end{equation}
where $\mathbb{E}_{V_{A B C}}\left[F_{B}\right]$ denotes the ensemble-averaged expectation value of $F_{B}$ after applying the glued random circuit $V_{A B C}$ to the initialized state. This remains parametrically separated from the gauge-symmetric Haar value $\mathbb{E}_{U_{A B C}}\left[F_{B}\right]=\frac{E_{\max }}{2}$. Thus, measuring $F_{B}$ efficiently distinguishes the glued ensemble from gauge-symmetric Haar randomness.

This distinguishing signal disappears once the patches exceed the diffusive scale: $\xi(n) \gg c^2(n)$. In this regime, the dimensions of the boundary-flux sectors become equal at leading order, so the initial boundary condition no longer produces a parametrically detectable bias at the opposite edge:
\begin{equation} \label{eq:ExpecValueE01largeMain}
\mathbb{E}_{V_{ABC}}[F_{B}] \approx \frac{E_{\max}}{2} = \mathbb{E}_{U_{ABC}}[F_{B}]~.
\end{equation}
The detailed derivations of Eqn.~\eqref{eq:ExpecValueE01SmallMain} and Eqn.~\eqref{eq:ExpecValueE01largeMain} are given in Supplementary Material, Section~II.A. Thus the diffusive scale marks the characteristic patch size below which residual boundary-flux memory obstructs Haar-like gluing. 

\begin{theorem}[No-go for subdiffusive gluing of gauge-symmetric random unitaries]
\label{thm:NoGoSublinearmain}
For a \(\mathrm{U}(1)\) lattice gauge theory, glued circuits built from physical gauge-symmetric random unitaries supported on overlapping subdiffusive patches cannot reproduce gauge-symmetric Haar randomness on \(\mathcal{H}_{\mathrm{phys}}\).
\end{theorem}

Gauge symmetry does not rule out efficient formation of Haar-like random dynamics altogether, but instead introduces a characteristic size that the support of constituent random unitaries must exceed. This places gauge-symmetric random dynamics in an interesting intermediate regime between globally symmetric and unconstrained systems. Unlike a global conservation law, Gauss's law does not enforce global correlations through branching rules  \cite{Marvian_2022,Marvian_2024,Li_2024,Li_2025,li2024efficientquantumpseudorandomnessconservation,Zheng:2021sml}; yet unlike an unconstrained system, it generates nontrivial correlations between local flux sectors that persist up to the diffusive scale. Gauge-symmetric systems therefore provide a natural setting in which the crossover from symmetry-constrained to effectively unconstrained randomization can be studied quantitatively. Physically, this crossover is controlled by how efficiently local matter charges can screen boundary-flux mismatches: as shown in Theorem~\ref{thm:LocalRandomization}, enriching the matter-charge spectrum increases the number of allowed physical configurations within a local region, thereby shortening both the randomization and diffusive scales and allowing Haar-like behavior to emerge on smaller patches. The same argument extends to general Abelian lattice gauge theories through the corresponding definition of the diffusive scale; see Supplementary Material, Section~II.A.

\subsection{Gauge-symmetric approximate random unitaries in logarithmic-depth}

Theorem~\ref{thm:NoGoSublinearmain} leaves open the possibility of shallow gauge-symmetric Haar-like dynamics once the local patches exceed the diffusive scale. We now prove a complementary gauge-symmetric gluing theorem directly on $\mathcal H_{\mathrm{phys}}$ and use it to construct low-depth approximate unitary $k$-designs by gluing gauge-symmetric random unitaries on overlapping patches, achieving logarithmic depth for systems with $\mathrm{diff}(n)=\mathcal O(1)$. When gauge charges can be completely screened locally, this construction further yields logarithmic-depth approximate strong designs and strong PRUs. These provide two complementary notions of approximation to the physical gauge-symmetric Haar ensemble: unitary $k$-designs reproduce its moments up to order $k$, whereas PRUs require computational indistinguishability from it by efficient quantum algorithms.

Quantitatively, a random unitary ensemble \(\mathcal E\) forms an \(\varepsilon\)-approximate gauge-symmetric unitary \(k\)-design if
\begin{equation}
\label{eq:ApproxPhysGaugeSymDesign}
(1-\varepsilon)\Phi_{H}^{\boldsymbol{\mathrm{U}(1)}} \preceq \Phi_{\mathcal E}^{\boldsymbol{\mathrm{U}(1)}} \preceq (1+\varepsilon)\Phi_{H}^{\boldsymbol{\mathrm{U}(1)}}~,
\end{equation}
where \(\Phi_1\preceq\Phi_2\) means that \(\Phi_2-\Phi_1\) is completely positive. The gauge-symmetric Haar twirl is defined as
\begin{equation}
\label{eq:PhysGaugeSymHaarTwirl}
\Phi_{H}^{\boldsymbol{\mathrm{U}(1)}}(\rho_{\text{phys}}) \equiv \mathbb E_{U\sim H_{\boldsymbol{\mathrm{U}(1)}}^{\text{phys}}} \left[ U^{\otimes k} \rho_{\text{phys}} U^{\dagger\otimes k} \right]~,
\end{equation}
where $H^{\text{phys}}_{\boldsymbol{\mathrm{U}(1)}}$ denotes the Haar measure on $\mathrm{U}(D_{\text{phys}})$ acting on the physical Hilbert space $\mathcal H_{\mathrm{phys}}$. The ensemble twirl $\Phi_{\mathcal E}^{\boldsymbol{\mathrm{U}(1)}}$ is defined analogously, and $\varepsilon$ specifies the relative approximation error~\cite{schuster2025randomunitariesextremelylow,schuster2025strongrandomunitariesfast,Brand_o_2016}. By contrast, a physical gauge-symmetric pseudorandom unitary (PRU) ensemble is computationally indistinguishable from \(H_{\boldsymbol{\mathrm{U}(1)}}^{\text{phys}}\) to any efficient quantum algorithm with oracle access to the sampled unitary. Thus, unitary designs quantify closeness through moments, whereas PRUs quantify it through efficient distinguishability.

\begin{figure}[!ht] 
\centering
\includegraphics[width=0.3\textwidth]{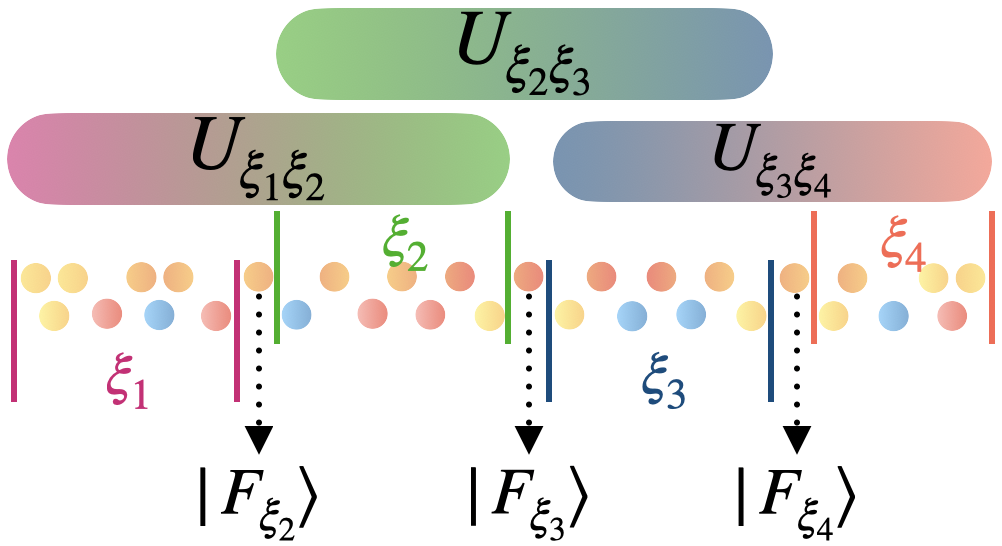}
\caption{The staggered two-layer circuit has slightly redefined local patches with all controlled unitaries within each layer acting in parallel. Gauge-symmetric unitaries $U_{\xi_1\xi_2}$ and $U_{\xi_3\xi_4}$ in the first layer are both controlled the on odd interface flux state $\ket{F_{\xi_3}}$. Whereas the controlled flux is randomized by $U_{\xi_2\xi_3}$ in the second layer which is controlled on even interface flux states $\ket{F_{\xi_2}}$ and $\ket{F_{\xi_4}}$.}
\centering \label{Fig:ParallelControlMain}
\end{figure}

The main technical ingredient is a gauge-symmetric analogue of the gluing constructions for unconstrained and symmetry-constrained random unitaries~\cite{schuster2025randomunitariesextremelylow,schuster2025hardnessrecognizingphasesmatter}. We again divide the chain into $m$ consecutive patches $\xi_1,\ldots,\xi_m$, separated by individual gauge links carrying interface fluxes $F_{\xi_2},\ldots,F_{\xi_m}$ (Figure~\ref{Fig:ParallelControlMain}). Each local gauge-symmetric unitary $U_{\xi_j\xi_{j+1}}$ acts on two neighboring patches together with their shared interface link, with joint support: $\widehat{\xi}_{j,j+1} = \xi_j\cup\ket{F_{\xi_{j+1}}}\cup\xi_{j+1}~.$ The internal interface flux is randomized, while the exterior boundary fluxes act as controls and remain unchanged. Independently sampled controlled gauge-symmetric unitaries can therefore be arranged into a two-layer brickwork architecture. At sufficiently large length scales, the approximate factorization of physical-sector dimensions allows these local random unitaries to be glued into a global approximate design with controlled error.

\begin{theorem}[Gluing gauge-symmetric random unitaries]
\label{thm:GluingThmMain}
Consider a $\mathrm{U}(1)$ lattice gauge theory on a one-dimensional open chain with $n$ sites, matter charges $q_i\in\{-1,0,1\}$, and gauge-flux spectrum $E_{i,i+1}\in\{0,1,\ldots,E_{\max}(n)\}$. Partition the chain into consecutive patches of equal size $\xi(n)$. For target error $0<\varepsilon\leq 1$, consider a two-layer brickwork ensemble $\mathcal E$ whose local unitaries are independently sampled from $(\varepsilon/n)$-approximate controlled physical gauge-symmetric unitary $k$-designs acting on neighboring patch pairs. Then $\mathcal E$ forms an $\varepsilon$-approximate physical gauge-symmetric unitary $k$-design on $\mathcal H_{\mathrm{phys}}$, provided
\begin{equation}
\label{eq:GluePatchSizeLowerBoundMain}
\xi(n) \gg \mathrm{diff}(n)\log\left(\frac{nk^2 \cdot c(n)^{\mathcal{O}(k)}}{\varepsilon\,\mathrm{diff}(n)}\right)~.
\end{equation}
\end{theorem}

The patch-size condition in Eqn.~\eqref{eq:GluePatchSizeLowerBoundMain} reflects two sources of gluing error. First, the approximate factorization in Eqn.~\eqref{eq:Factorization} exponentially suppresses residual boundary-flux correlations on the diffusive scale. Second, higher moments involve additional contributions from different permutations of the $k$ copies, which are suppressed by the large dimensions of the overlapping physical sectors but carry gauge-sector prefactors bounded by $c(n)^{\mathcal O(k)}$. Thus, although both contributions decay rapidly with increasing patch size, the latter prefactors become increasingly restrictive as $k$ grows for $c(n)>1$. In particular, as shown in Eqn.~\eqref{eq:GluePatchSizeLowerBoundMain}, our construction cannot accommodate design orders $k=\Theta(n)$ with sublinear patch sizes $\xi(n)=o(n)$. This limitation also prevents a direct extension of the PRU gluing argument of Ref.~\cite{schuster2025randomunitariesextremelylow} to our construction for $c(n)>1$. That argument requires control at superpolynomial design orders to establish security against all polynomial-time quantum adversaries, whereas our present bounds do not support such orders with sublinear patch sizes.

The case $c(n)=1$ is qualitatively different: as discussed above, the physical Hilbert space becomes isomorphic to an unconstrained tensor-product space, and the additional gauge-sector corrections vanish exactly. Consequently, the gluing construction recovers the unconstrained result~\cite{schuster2025randomunitariesextremelylow}, allowing superpolynomially growing design orders $k=n^{\omega(1)}$ with approximation errors $\varepsilon=n^{-\omega(1)}$ for suitably chosen patch sizes $\xi(n)=\omega(\log n)$. Our construction builds on the gluing framework of Refs.~\cite{schuster2025randomunitariesextremelylow,schuster2025hardnessrecognizingphasesmatter}, with additional modifications to account for the Gauss-law constraints and boundary-flux-sector corrections. The full proof and more explicit constructions are given in Supplementary Material, Section~III.B.

Theorem~\ref{thm:GluingThmMain} provides a route to constructing low-depth physical gauge-symmetric unitary designs from local random unitaries. For strictly locally random systems with $c(n)=1$, the construction of physical gauge-symmetric PRUs becomes particularly simple. As established in Theorem~\ref{thm:LocalRandomization}, $\mathcal H_{\mathrm{phys}}$ is isomorphic to an unconstrained tensor-product Hilbert space of gauge-link degrees of freedom, each with local dimension $D_i=E_{\max}+1$. Local operations on these effective qubits can be implemented efficiently while preserving Gauss's law, allowing the pseudorandom-function components of the Luby--Rackoff--Function--Clifford (LRFC) construction to be realized on the physical Hilbert space~\cite{schuster2025strongrandomunitariesfast,schuster2025hardnessrecognizingphasesmatter,cui2025randomunitarieshamiltoniandynamics}. Moreover, when $D_i=2$ is prime, the corresponding Clifford ensemble forms a unitary $2$-design, providing the remaining ingredient required by LRFC. Hence, under the same computational assumptions as in the unconstrained setting, the LRFC construction can be implemented efficiently on $\mathcal H_{\mathrm{phys}}$, yielding low-depth physical gauge-symmetric PRUs with the pseudorandomness guarantees of Refs.~\cite{schuster2025strongrandomunitariesfast,schuster2025hardnessrecognizingphasesmatter}.

\begin{corollary}[Low-depth physical gauge-symmetric designs and PRUs] \label{cor:logDepthUnitaryDesignPRU}
Consider a strictly locally random \(\mathrm{U}(1)\) lattice gauge system on a 1D open chain, with \(E_{\max}=1\). Then 1D gauge-symmetric random circuits on \(\mathcal{H}_{\text{phys}}\) form an \(\varepsilon\)-approximate unitary \(k\)-design in circuit depth $d=\mathcal{O}\left(\log(n/\varepsilon)\cdot k\operatorname{poly}(\log k)\right)$~\cite{metger2024simple,schuster2025randomunitariesextremelylow}. Moreover, assuming that the Learning With Errors problem cannot be solved by subexponential-time quantum algorithms \cite{regev2024latticeslearningerrorsrandom,gheorghiu2024estimatingentropyshallowcircuit,Wang:2025gji}, there exist physical gauge-symmetric PRU ensembles, secure against polynomial-time quantum adversaries, that can be implemented by all-to-all gauge-symmetric circuits of depth $d=\operatorname{poly}(\log\log n)$~\cite{schuster2025hardnessrecognizingphasesmatter}.
\end{corollary}
Here, all-to-all connectivity allows arbitrary two-qubit gates between any pair of effective gauge-link qubits in the tensor-product representation of \(\mathcal H_{\mathrm{phys}}\). Let \(u_{j,\ell}\) denote such a gate acting on distinct link qubits \(j\) and \(\ell\), with identity action on all other effective qubits. Writing \(\mathcal W\) for the unitary isomorphism from the effective tensor-product space to \(\mathcal H_{\mathrm{phys}}\), we denote the corresponding gauge-invariant operation in the original link–matter description by \(\widetilde u_{j,\ell}\), satisfying $\widetilde u_{j,\ell} =\mathcal W u_{j,\ell}\mathcal W^\dagger~.$ The lifted gates act on the selected gauge links and their adjacent matter sites, with the matter charges updated coherently to preserve Gauss’s law. Thus, the all-to-all architecture relaxes geometric locality without modifying the underlying one-dimensional gauge constraints. Further implementation details are given in Supplementary Material, Section~III.C.2.

The logarithmic-depth design above is a particularly simple representative example. More generally, the gluing construction yields polylogarithmic-depth physical gauge-symmetric unitary designs whenever the randomization scale is bounded, $c(n)=\mathcal O(1)$, corresponding to a constant diffusive scale $\mathrm{diff}(n)=\mathcal O(1)$ in the one-dimensional model considered here. In this regime, $\varepsilon$-approximate unitary $k$-designs can be constructed with circuit depth $d=\mathcal O\left(\operatorname{poly}\log(n/\varepsilon)\cdot\operatorname{poly}(k)\right)$ on any circuit geometry. The full proofs, implementation details of the all-to-all gauge-symmetric circuits, and broader circuit-geometry-dependent constructions are given in Supplementary Material, Sections~III.B.3 and III.C.2.

\begin{figure*}[!ht]
\centering
\includegraphics[width=\linewidth]{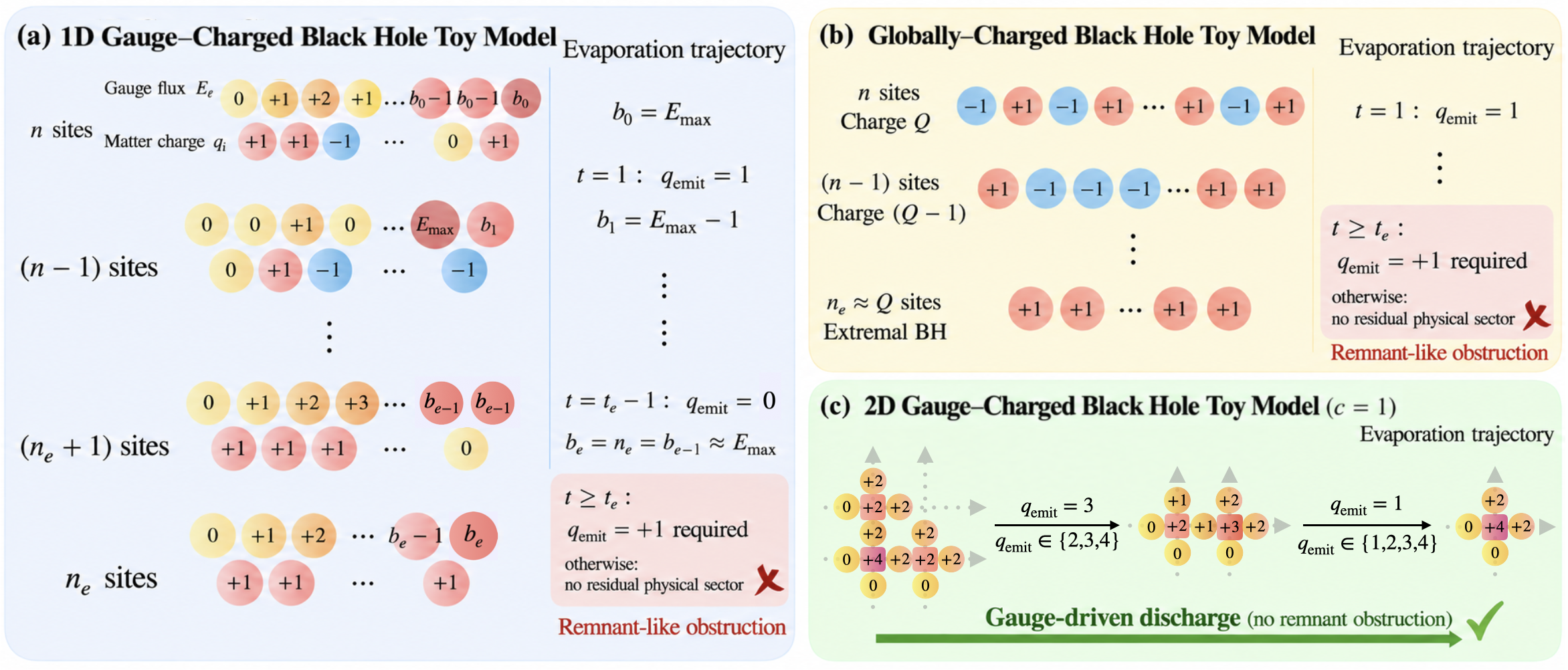}
\caption{(a) One-dimensional gauge-charge evaporation toy model with $E_{\ell} \in\left\{0, \ldots, E_{\max }\right\}$ and $q_i \in\{-1,0,1\}$. Matter charges $q_i=+1,0,-1$ are shown in red, yellow, and blue, respectively, while darker red gauge variables denote larger electric flux. Evaporation can drive the system toward an extremal configuration at $t_e$, after which a complete uninterrupted discharge trajectory is exponentially suppressed in the randomization scale. (b) Global-charge analogue with approximately charge-blind emission. Red and blue sites carry charges $+1$ and $-1$. A black hole with initial charge $Q$ can shrink without efficiently discharging until it reaches an extremal configuration with $n_e \simeq Q$, where continued evaporation requires a highly constrained sequence of charge-reducing emissions and is exponentially suppressed in $n_e$. (c) Two-dimensional gauge-charge example with $E_{\ell} \in\{0,1,2\}$ and $q_i \in\{-4,-3, \ldots, 4\}$. Matter charges are shown as red squares and gauge-link variables as yellow-to-orange circles, with darker colors indicating larger matter charge and electric flux, respectively. As the horizon shrinks, the flux it can support decreases, so the locally constrained charge-blind rule naturally biases the admissible emissions toward reducing the black-hole charge. Because the system is strictly locally random $(c=1)$, every locally admissible emission has a compatible residual physical sector, and evaporation proceeds without the remnant-like obstruction.}
\label{fig:Remnant}
\centering
\end{figure*}

\subsection{Efficient ``black hole'' discharge and gauge-symmetric fast scrambling through complete local screening}
\label{subsec:FastScramblingChargedBH}

The randomization scales introduced in Theorem~\ref{thm:LocalRandomization} suggest a possible connection between gauge-symmetric information scrambling and charged black-hole dynamics in quantum gravity. The mild electric Weak Gravity Conjecture (WGC) is motivated in part by the expectation that charged black holes possess sufficiently light charged states through which they can discharge, thereby avoiding stable extremal remnants~\cite{Arkani_Hamed_2007,Harlow_2023,Cheung_2018,fisher2017semiclassicalentropicproofweak,Hod_2017}. To explore a lattice analogue of this intuition, we adapt the standard qubit model mimicking black-hole evaporation~\cite{avery2012qubitmodelsblackhole,HartmanQuantumGravityBlackHoles2015,Harlow2016Jerusalem} to include conserved global and gauge charges. In our coarse-grained version, the shrinking black-hole Hilbert space is repeatedly scrambled and then loses one charged degree of freedom at each evaporation step.

We begin with a hypothetical ``black hole'' carrying an exact global \(\mathrm U(1)\) charge. (See Figure~\ref{fig:Remnant}(b).) We model its horizon degrees of freedom by a chain of $n$ effective sites with $q_i=\pm1$ and fixed total charge $Q$. The charge is distributed across these sites, which collectively represent the effective horizon rather than placing it at either endpoint of the chain. At each evaporation step, as in the unconstrained qubit model \cite{avery2012qubitmodelsblackhole}, its state is first Haar-randomized within the fixed-charge sector \(\mathcal H_{n,Q}\). Physically, semiclassical Hawking radiation is generated by local quantum fields near the horizon rather than by ejecting a charged constituent already stored inside the black hole. For an exact global symmetry, there is no associated long-range gauge field that can bias this local emission process toward the sign of charge required to discharge the black hole, reflecting the approximately thermal character of the ``Hawking radiation''~\cite{Hawking1975,Harlow_2019,harlow2019symmetriesquantumfieldtheory,Harlow_2023}. Therefore, in the second stage of each evaporation step, the environment proposes an emitted charge \(q_{\mathrm{emit}}=\pm1\) with equal probability, independently of the current value of \(Q\). We refer to this idealized infinite-temperature rule as \emph{charge-blind emission}. The proposal is physically allowed only if the residual sector \(\mathcal H_{n-1,Q-q_{\mathrm{emit}}}\) is nonempty. If it is empty, we terminate that evaporation history rather than resampling the emitted charge. Resampling until an allowed charge is obtained would condition the emission on the black-hole charge and thereby introduce an effective discharge mechanism into the otherwise charge-blind process.

We follow the semiclassical evaporation only down to a one-site configuration, representing the smallest black hole retained within our coarse-grained effective description; its subsequent evolution lies in the genuinely quantum-gravitational regime and is not modeled here \cite{Hawking1975}. Because the system size decreases while the charge changes only diffusively, the black hole typically approaches an extremal configuration with \(n_e\approx |Q|\) remaining sites carrying charge of the same sign. From this point, a complete physical evaporation trajectory under the charge-blind proposal process requires $n_e-1$ consecutive emissions with $q_{\text {emit }}=\operatorname{sgn}(Q)$. Under the charge-blind proposal process, such a trajectory occurs with probability $P_{\text {discharge }}=2^{-\left(n_e-1\right)}=\exp \left[-\Theta\left(|Q|\right)\right]$, so complete late-time discharge becomes exponentially atypical as the residual extremal charge grows. This reproduces, within the toy model, the familiar remnant intuition associated with exact global symmetries. Indeed, avoiding this late-time bottleneck by allowing the residual charge to be compressed into progressively fewer sites would require an unbounded on-site charge spectrum, so that the one-site endpoint could carry arbitrarily large global charge, in tension with the Bekenstein-Hawking entropy formula and possibly become an infinite family of remnants ~\cite{Harlow_2023}. We therefore refer to the exponentially suppressed discharge at finite charge capacity as a \emph{remnant-like obstruction}. This provides a quantum-information analogue of the black-hole remnant argument against exact global symmetries in quantum gravity~\cite{Kallosh:1995hi,harlow2019symmetriesquantumfieldtheory}.

We next apply the same proposal-and-projection logic to a gauge-charged black hole. (See Figure~\ref{fig:Remnant}(a).) Its charged degrees of freedom are represented by a shrinking lattice gauge system, with the one-site configuration denoting the smallest black hole retained within the semiclassical effective description. Site removal models the transfer of an effective charge-carrying degree of freedom to the radiation. The coarse-grained prescription specifies emission channels through their charges and Gauss-law constraints, isolating the kinematic screening obstruction to discharge. At evaporation step $t$, the black hole occupies the physical sector $\mathcal{H}_{n_t}\left(a, b_t\right)$, with $n_t$ remaining sites and incoming and outgoing boundary fluxes $a, b_t \in\left\{0,1, \ldots, E_{\max }\right\}$. For simplicity, we identify the outgoing (right) boundary of the chain with the effective horizon through which Hawking radiation is emitted, so $a$ is held fixed while $b_t$ evolves with the emitted charge. Internal scrambling is modeled by a Haar-random physical unitary $U_t \in \mathrm{U}\left(D_{n_t}\left(a, b_t\right)\right)$, which preserves the boundary fluxes and hence the total black-hole charge during the scrambling step.

Charge exchange with the exterior occurs only in the subsequent emission step. Emitting a matter excitation of charge \(q_t\) exposes a new boundary flux \(b_{t+1}=b_t-q_t\) and proposes the transition
\begin{equation}
\mathcal H_{n_t}(a,b_t) \longrightarrow \mathcal H_{n_t-1}(a,b_t-q_t)\otimes\mathcal R_{q_t}~,
\end{equation}
where $\mathcal{R}_{q_t}$ denotes the radiation sector carrying the emitted charge $q_t$. As in the global-charge model, we take the underlying Hawking proposal to be otherwise charge blind. Unlike a global charge, however, gauge charge is accompanied by electric flux through the boundary, so the emission must respect the local Gauss-law kinematics of the newly exposed horizon. We therefore sample \(q_t\) uniformly from the locally admissible set
\begin{equation}
    \mathcal Q_{b_t} = \left\{ q_t\in\{-q_{\max},\ldots,q_{\max}\}:~0\leq b_{t+1}\leq E_{\max} \right\}~.
\end{equation}
We then ask whether the corresponding smaller physical sector exists, \(D_{n_t-1}(a,b_t-q_t)>0\). If it does, the emission is allowed and the black hole evolves to \(\mathcal H_{n_t-1}(a,b_t-q_t)\); otherwise the proposal is rejected. We refer to this as the \textit{locally-constrained charge-blind emission}.

The restriction \(q_t\in\mathcal Q_{b_t}\) removes purely local violations of the electric-flux cutoff, so any remaining rejection probes a distinct many-body obstruction: although the emission is locally compatible with the gauge field, the available matter spectrum may still be unable to realize the required residual boundary-flux sector. In contrast to the global-charge toy model, the gauge field itself carries information about the black-hole charge through the boundary flux. Consequently, the locally admissible emission channels depend on the flux supported by the newly exposed horizon and can become naturally biased toward discharge as the horizon shrinks. This provides a lattice analogue of the familiar discharge mechanism of continuum charged black holes, where the exterior gauge field and horizon electrostatic potential bias charged Hawking emission and Schwinger pair production toward reducing the black-hole charge \cite{harlow2019symmetriesquantumfieldtheory,Harlow_2019}. In the one dimensional toy model (Figure~\ref{fig:Remnant}(a)), this tendency is not fully manifest, because the effective horizon is zero-dimensional and does not shrink with the black-hole region; this is an artifact of the simplified geometry. The corresponding gauge-driven discharge is illustrated explicitly in the two-dimensional example shown in Figure~\ref{fig:Remnant}(c). Nevertheless, the one-dimensional model retains the essential screening question—whether the available charged matter can realize the sequence of boundary-flux transitions required for complete discharge. We therefore focus primarily on the one-dimensional model, while using the two-dimensional example to illustrate the additional effect of a shrinking horizon.

Consider the usual one-dimensional \(\mathrm U(1)\) model with \(E_\ell\in\{0,\ldots,E_{\max}\}\) and \(q_i\in\{-1,0,1\}\), for which \(c=E_{\max}\). (See Figure~\ref{fig:Remnant}(a).) Starting from boundary fluxes \(a=0\) and \(b_0=E_{\max}\), charge-blind evaporation can drive the shrinking black hole toward an extremal \(n_e\)-site configuration in which every remaining site carries \(q_i=+1\). Let \(b_e\) denote the outgoing flux at this stage. Gauss's law gives \(n_e=b_e\); because the charge changes only diffusively, up to the slight bias near the flux cutoff, \(b_e\simeq E_{\max}\), and hence \(n_e\simeq E_{\max}=c\). From this point onward, only emissions with \(q_{\text{emit}}=+1\) lead to nonempty physical sectors of the successively smaller extremal black hole. The extremal tail therefore contains \(\mathcal O(c)\) constrained evaporation steps, at each of which only one of the locally-admissible charge-emission proposals continues the discharge, with probability approximately \(1/3\).

Consequently, the probability for the locally constrained charge-blind process to realize a complete physical discharge trajectory is exponentially small, \(P_{\mathrm{discharge}}=\exp[-\Theta(c)]\). The randomization scale thus acquires a direct physical interpretation in the evaporation problem: the same restricted matter spectrum that increases the spatial extent required to screen an admissible flux mismatch also produces an \(\mathcal O(c)\)-long constrained extremal-discharge tail. The resulting exponential suppression of complete late-time discharge is the gauge-theoretic analogue of the remnant-like obstruction identified above for globally symmetric evaporation.

An analogous obstruction persists in toy models on higher-dimensional lattices, where the gauge-driven tendency toward discharge becomes more transparent. For instance, in our two-dimensional square-lattice toy model, we identify the effective horizon with the boundary links connecting the remaining black-hole region to its exterior, regardless of their flux orientations. At each evaporation step, we select a boundary site with the maximal number of such links, breaking ties uniformly at random. We then apply the same locally constrained, charge-blind evaporation rule as in one dimension: the emitted charge is sampled from the locally admissible set, and the process terminates if the resulting physical Hilbert-space sector has zero dimension. This prescription determines the evaporation trajectory dynamically, without specifying a path in advance. Since the selected site has at least two exterior links, its removal cannot increase the number of boundary links forming the effective horizon. Consequently, as the black-hole region shrinks, the maximal total gauge charge that can be supported cannot increase, naturally favoring discharge. Nevertheless, if the available matter spectrum cannot efficiently screen the residual boundary flux, the system can still enter an extremal regime in which further evaporation requires a highly restricted sequence of charged emissions. Repeated charge-blind sampling then makes complete discharge exponentially unlikely, giving rise to the same remnant-like obstruction as in one dimension. Explicit two-dimensional constructions are given in Supplementary Material, Section~IV.A.2.

The remnant-like obstruction disappears in the strictly locally random regime of Theorem~\ref{thm:LocalRandomization}. When \(c=1\), every admissible boundary-flux mismatch can be screened within a single site, so locally constrained charge-blind evaporation no longer relies on an exponentially unlikely sequence of screening steps. In the one-dimensional model, this requires the available matter spectrum to contain every charge \(q\in\{-E_{\max},\ldots,E_{\max}\}\). The physical significance of local screening is also clear in our two-dimensional square-lattice toy model: as the black-hole region shrinks, the number of horizon links and hence the maximal supported gauge charge decrease, naturally driving the system toward discharge. When the available matter charges can sufficiently screen the residual boundary flux, evaporation can proceed without encountering the remnant-like obstruction, as illustrated by the successful discharge trajectory in Figure~\ref{fig:Remnant}(c). For physical discharge, the relevant charged channels must also be energetically accessible. This connects our screening requirement to the discharge principle motivating the Weak Gravity Conjecture~\cite{Arkani_Hamed_2007,Harlow_2023}. We therefore propose a lattice-screening refinement of this principle, combining complete local screening with an energetic condition on the channels required for extremal semiclassical discharge:

\begin{conjecture}[Lattice-Screening Weak Gravity Conjecture]
\label{conj:latticeWGCMain}
In a coarse-grained $\mathrm{U}(1)$ lattice description of a semiclassical black hole, the available matter spectrum supports a Gauss-law-consistent one-site physical state for every boundary-flux configuration permitted by the incident link Hilbert spaces. Moreover, every nonzero charge channel required for extremal semiclassical discharge is realized by an energetically accessible superextremal state, whose charge-to-mass ratio exceeds the extremal black-hole bound.
\end{conjecture}

Under the toy-model assumption that each allowed local charge labels a unique matter state, the screening content of Conjecture~\ref{conj:latticeWGCMain} is equivalent to strict local randomization, \(c=\mathrm{diff}=1\). Together with the superextremality requirement, this removes the kinematic and energetic obstructions to efficient discharge identified in our evaporation model. Semiclassically, gauge charge differs from global charge because its exterior electric field can bias charged emission toward discharge. Our model reflects this distinction through flux-dependent emission channels, while strict local randomization removes the remaining obstruction from insufficient matter-charge content. The role of charged matter suggests a qualitative parallel with charge completeness, formulated as Conjecture~2 in Ref.~\cite{harlow2019symmetriesquantumfieldtheory}, according to which bulk dynamical objects populate every finite-dimensional irreducible representation of the gauge group. The charged states supported on our effective lattice sites provide a heuristic counterpart of these bulk objects. For \(\mathrm U(1)\), a matter spectrum containing every integer charge would guarantee \(c=1\) for any finite electric-flux cutoff. This comparison concerns charge spectrum, while energetic accessibility of the channels needed for discharge enters separately in the proposed analogy. If the screening principle extends beyond the toy model, it may suggest constraints on the charged spectrum relevant to semiclassical discharge beyond superextremality alone~\cite{AalsmaColeShiu2019}.

Remarkably, the same screening condition also favors fast scrambling in our construction: complete local screening eliminates both the additional gauge-induced randomization scale and the gauge-sector corrections that become increasingly restrictive at higher design orders. Our construction therefore recovers the higher-order randomization guarantees of unconstrained systems, suggesting that gauge-invariant information can scramble just as rapidly. This raises the possibility that efficient discharge and fast scrambling of gauge-charged black holes are linked by a common requirement on the available charged spectrum. Motivated by the combined screening and energetic discharge requirements of Conjecture~\ref{conj:latticeWGCMain}, we propose the following gauge-charged extension of the fast-scrambling conjecture~\cite{Hayden_2007,Sekino_2008,schuster2025strongrandomunitariesfast}.

\begin{conjecture}[Gauge-charged fast scrambling conjecture]
\label{con:physicalgaugesymFastScram}
A gauge-charged black hole can scramble gauge-invariant information as rapidly as an uncharged black hole of the same entropy, provided that the underlying theory satisfies an appropriate form of the weak gravity conjecture, supplemented by a complete charge-screening requirement, so that efficient discharge is guaranteed. With all-to-all gauge-invariant interactions, the scrambling time can attain $t_{\mathrm{scr}}=\Theta(\log S)$, where $S$ is the black-hole entropy.
\end{conjecture}

Our lattice results provide operational evidence for Conjecture~\ref{con:physicalgaugesymFastScram}. As argued in ref.~\cite{schuster2025strongrandomunitariesfast}, fast scrambling is more naturally captured by strong unitary designs and strong PRUs, which remain pseudorandom under access to \(U\), \(U^\dagger\), \(U^*\), and \(U^T\), and encompass diagnostics such as out-of-time-order correlations and Hayden--Preskill-type decoding~\cite{cotler2022information,yoshida2017efficient,Zhandry2025HowTM}. For a strictly locally random system, \(\mathcal H_{\mathrm{phys}}\cong(\mathbb C^{D_i})^{\otimes L}\), where \(D_i\) is the local gauge dimension and \(L\) is the number of links. The strong-randomness constructions of ref.~\cite{schuster2025strongrandomunitariesfast} can therefore be implemented directly on the physical gauge degrees of freedom and combined through our staggered two-layer gauge-symmetric gluing construction. Under all-to-all connectivity, this yields physical gauge-symmetric strong unitary designs and strong PRUs in \(\operatorname{poly}(\log n)\) depth for prime \(D_i\), with the strong-design depth sharpening to \(O(\log n)\) for \(D_i=2\), saturating the corresponding \(\Omega(\log n)\) scrambling lower bound. 

The same evaporation analogy and (strong-)unitary design/PRU constructions extend to finite Abelian gauge groups. Unlike the truncated \(\mathrm U(1)\) case, a complete local discrete-charge spectrum removes the screening obstruction and makes every global gauge-charge sector strictly locally random, so the logarithmic-depth constructions apply sector by sector; see Supplementary Material, Section V. Together, these results provide operational support for the gauge-charged fast-scrambling conjecture and suggest a possible connection between the lattice-screening weak gravity conjecture and fast scrambling: the same local charge-screening structure removes both the kinematic discharge obstruction in our toy models and the additional gauge-induced obstruction to rapid randomization in our circuit constructions.

\section{Discussion}

Gauge invariance places exact constraints on the physical Hilbert space, but these constraints need not impose equally strong restrictions on quantum dynamics at all length scales. Our results illustrate how the ability of local matter to screen gauge flux determines the extent to which distant regions remain correlated, and consequently how efficiently gauge-symmetric randomization can be achieved. In particular, the distinction between approximate screening over a finite length scale and complete screening within a single site becomes especially consequential for higher-order randomization. This suggests that the scrambling properties of gauge theories cannot be understood from symmetry alone, but depend sensitively on the microscopic spectrum of charged excitations.

The same perspective offers a complementary interpretation of the discharge problem for gauge-charged black holes. In our evaporation toy models, gauge-flux constraints can favor discharge as the black hole shrinks, while the available matter-charge spectrum determines whether evaporation can proceed without encountering a remnant-like obstruction. Although these simplified models do not establish a general criterion for black-hole discharge or scrambling, they highlight a possible common role for charge screening in both phenomena. Understanding this connection may clarify how quantum-gravitational consistency conditions constrain not only which charged states must exist, but also how efficiently gauge-invariant information can be redistributed.

\section*{{Acknowledgements}}\label{sec:ack}
We thank Robert Huang, Zi-Wen Liu, and Han Zheng for their helpful insights. Y.W.~acknowledges support from a QuiCS Lanczos Fellowship. Y.L.~, J.W.~, and J.L.~are supported in part by the University of Pittsburgh, School of Computing and Information, Department of Computer Science, Pitt Cyber, Pitt Momentum fund, PQI Community Collaboration Awards, John C. Mascaro Faculty Scholar in Sustainability, Switzerland NSF award 2000-1-243053, NSF award 2535915, 2610010, Thinking Machines Lab and Cisco Research. This research used resources of the Oak Ridge Leadership Computing Facility, which is a DOE Office of Science User Facility supported under Contract DE-AC05-00OR22725. Y.W.~and~A.V.G.~acknowledge support from the U.S.~Department of Energy, Office of Science, Accelerated Research in Quantum Computing, Fundamental Algorithmic Research toward Quantum Utility (FAR-Qu). Y.W.~and~A.V.G.~were also supported in part by NSF RQS (award No.~2553574), NSF STAQ program, DoE ASCR Quantum Testbed Pathfinder program (award No.~DE-SC0024220), ONR MURI,  AFOSR MURI, ARL (W911NF-24-2-0107), and NQVL:QSTD:Design:FTL. Y.W.~and~A.V.G.~also acknowledge support from the U.S.~Department of Energy, Office of Science, National Quantum Information Science Research Centers, Quantum Systems Accelerator (award No.~DE-SCL0000121).

\bibliographystyle{apsrev4-1}
\bibliography{cite}

\pagebreak
\clearpage
\foreach \x in {1,...,\the\pdflastximagepages}
{
	\clearpage
	\includepdf[pages={\x}]{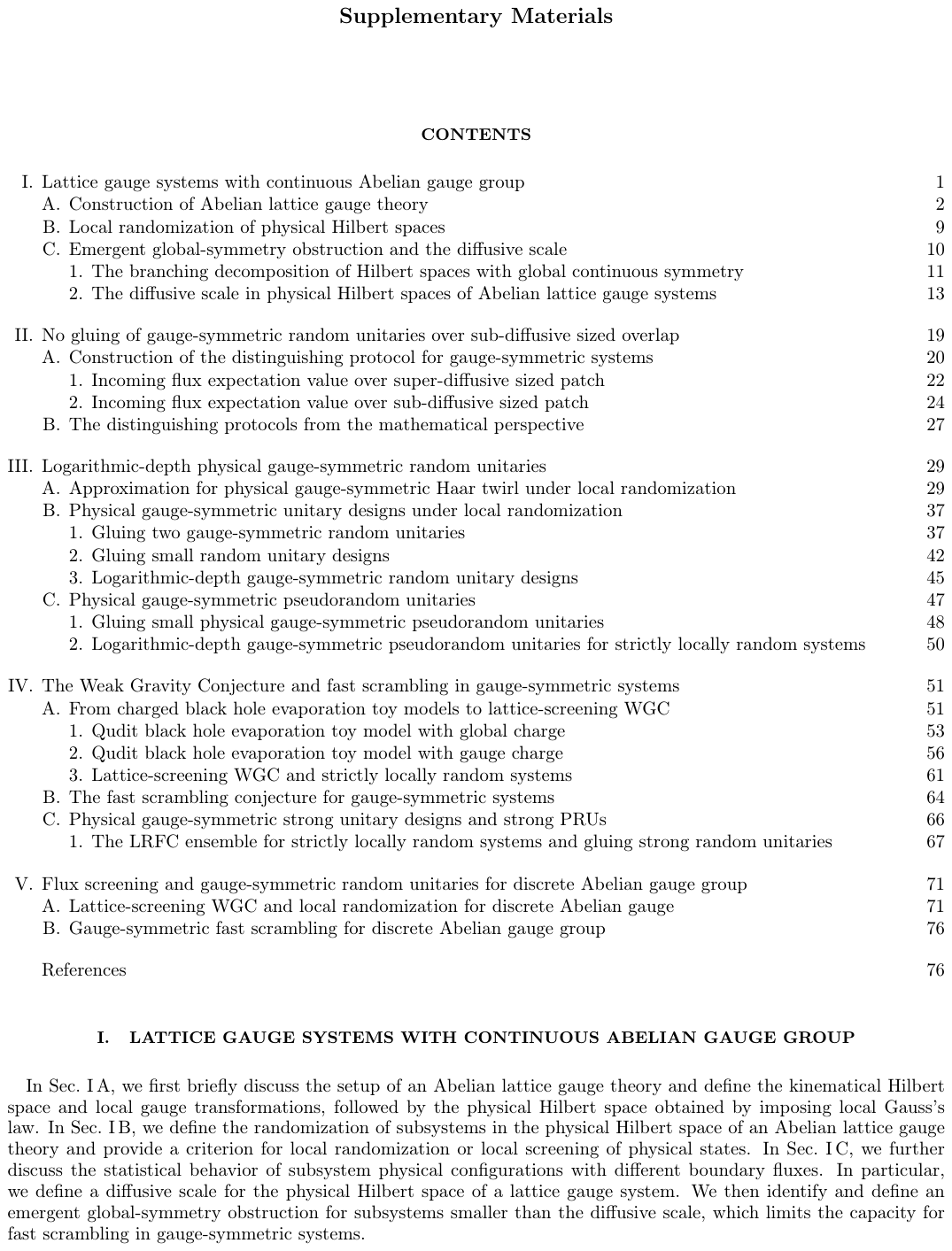}
}

\end{document}